\documentclass[aps,prl,twocolumn,groupedaddress]{revtex4}
\usepackage{graphicx}
\begin{document}

\title{Photoconduction and photocontrolled collective effects in the Peierls conductor TaS$_3$}
\author{S.~V. Zaitsev-Zotov}
\email[]{SerZZ@cplire.ru}
\author{V.~E. Minakova}
\affiliation{Institute of Radioengineering and Electronics,
Russian Academy of Sciences, Mokhovaya 11, bld.7, 125009 Moscow,
Russia}

\begin{abstract}
Light illumination of thin crystals of CDW conductor TaS$_3$ is
found to result in dramatic changes of both linear  ($G$) and
nonlinear conduction. The increase of $G$ is accompanied  by suppression of the collective 
conduction, growth of the threshold field $E_T$,  and appearance of the switching and
hysteretic behavior in the nonlinear conduction. 
The effects in the nonlinear conduction are associated with increase of CDW elasticity
due to illumination that leads in particular to appearance of a relation 
$E_T\propto G^{1/3}$ expected for the one-dimensional pinning. \\
{\em Published in:} Pis'ma v ZhETF, {\bf 79}, 680 (2004) [JETP Letters, {\bf 79}, 550 (2004)]

\end{abstract}


\maketitle

Quasi-one-dimensional (quasi-1D) conductors with charge-density-waves (CDW) \cite{Review} are
one of the most interesting physical systems with collective electron transport.
The interaction between electrons condensed into the CDW dominates in elastic properties 
of the electron crystal --- CDW. The elastic properties of the CDW affect
such characteristics of quasi-1D conductors as the value
of the threshold field for CDW sliding, $E_T$, phase-correlation length, dielectric constant, {\em et al}. 
In its turn the elastic properties are controlled by quasiparticles (electrons and holes) 
thermally excited over the Peierls gap in the energy spectrum and screening the electric
fields caused by CDW deformations. 
Thus a variation of the quasiparticle concentration (or the total carrier concentration) 
may be a tool controlling the properties of CDW conductors. An attempt to vary 
the total carrier concentration has been undertaken 
in the field effect experiment \cite{Adelman}. In particular it was found that
1\%-change of the total concentration of the current carriers by the transverse 
electric field leads to 40\%-change of the threshold 
field value. Another well-known way to modify the carrier concentration  
is excitation of nonequilibrium current carriers by light. 
For example, illumination of a semiconductor may result in increase of the 
carrier concentration by orders of magnitude. Such a change can be easily detected as 
a variation of the conduction (photoconduction). Photoconduction is one of the 
most fruitful methods to study the details of the energy structure, current carrier recombination time, 
and other semiconductor parameters. The similarity of numerous 
physical properties of the CDW conductors and semiconductors arising 
from the existence of the gap in the electron state density is well known 
\cite{Artemenko}.

Several attempts of experimental search for  photoconduction of CDW 
materials \cite{Brill,Itkis,Gaal,Ogawa} reveal contradictory results. In Ref.~\cite{Brill,Itkis} 
no noticeable photoconduction in TaS$_3$ was observed. Instead, the 
bolometric response was found and employed for detailed study of the 
energy structure in TaS$_3$. In addition an enhancement of the bolometric 
response was reported in nonlinear regime Ref.~\cite{Brill}.
In Ref.~\cite{Gaal} photoinduced CDW conduction 
was observed in blue bronze 
K$_{0.3}$MoO$_3$. The red boundary of the effect was found 
to correspond to the Peierls gap value. 
The phenomenon was associated 
with initiation of the CDW depinning by optically excited 
single electrons. 
No light-induced variation of the linear conduction and 
the threshold field was reported. 
In Ref.~\cite{Ogawa} photoinduced modification
of the dynamic transition from slide to creep in K$_{0.3}$MoO$_3$ 
was reported: light illumination was found to increase $E_T$ 
and the CDW creep rate. 
The origin of the effect was attributed to a local destruction 
of the CDW which led to the photoinduced phase slip and the 
redistribution of the CDW phase. 
No effect of illumination on the linear conduction was reported.
Thus, despite some similarity between
CDW conductors and semiconductors
no photoresponse in the linear conduction was found
during 25 years of study of the CDW materials.
Its worth to mention that the absence of photoconduction 
would agree with theories predicting very small quasiparticle 
lifetime, of the order of $10^{-12}$~s \cite{Brazovskii}	Femtosecond
spectroscopy study of K$_{0.3}$MoO$_3$ has shown 
that the electron-hole recombination time is short indeed, of 
subpicosecond scale \cite{Demsar}. 
From this point of view quasiparticles (electrons and holes) 
are ill-defined physical objects and the absence of photoconduction is 
a feature of CDW conductors. 
So observation of  photoconduction is essential 
for physics of CDW conductors.

Here we show that photoconduction of quasi-1D conductor TaS$_3$ 
can be directly observed in low-frequency conduction measurements. 
In particular, we show that the linear conduction 
may be increased up to an order of magnitude under the light illumination.
The resulting changes of the nonlinear conduction are dramatic,
and reveal themselves in a substantial growth of the threshold field, 
suppression of the nonlinear conduction near $E_T$,
and appearance of the switching behavior. We also show that the observed
growth of $E_T$ can entirely accounted by photoinduced increase of the 
CDW elasticity. Thus, no exotic assumptions on photoinduced phase slip  \cite{Ogawa} 
or CDW depinning  \cite{Gaal} is required. Moreover, our results 
clearly demonstrate the opposite effect, suppression of the 
collective conduction under light illumination.

Orthorhombic TaS$_3$ is a typical Peierls conductor. 
In this material the CDW formation at $T_P = 220$~K 
is accompanied by the complete dielectrization of the electron spectrum. All five 
studied samples (made by splitting of high-quality crystals) having 
cross-section areas $0.002\leq s\leq 0.15\;\mu$m$^2$ demonstrated 
qualitatively similar behavior. 
Such thin crystals were chosen to enhance the contribution of the region,  
affected by the illumination \cite{Comment1}.  
In our samples the photoconduction developed in the entire sample, rather 
than in a surface layer shunted by bulk like in Refs.~\cite{Brill,Itkis,Gaal,Ogawa}. 
Using of thin samples allows also to suppress heating effect due to 
exceptionally good thermal contact with sapphire substrate. In addition,
growth of potential relief due to finite-size effects \cite{Finite-Size} leads to
substantial  photocondution growth  due to spatial separation of photoexcited
electrons and holes. Current terminals were made by indium cold soldering, 
all measurements were performed in the two-terminal regime. IR LED with intensities 
$(10^{-6}$ - 1) W/cm$^2$ at the sample position, and with a wavelength of
$\lambda = 0.94$~$\mu$m was used. Thus the photon energy was higher than 
the optical gap value 125 meV  \cite{Itkis}.

\begin{figure}
 \vskip -2.3cm
 \includegraphics[scale=.45]{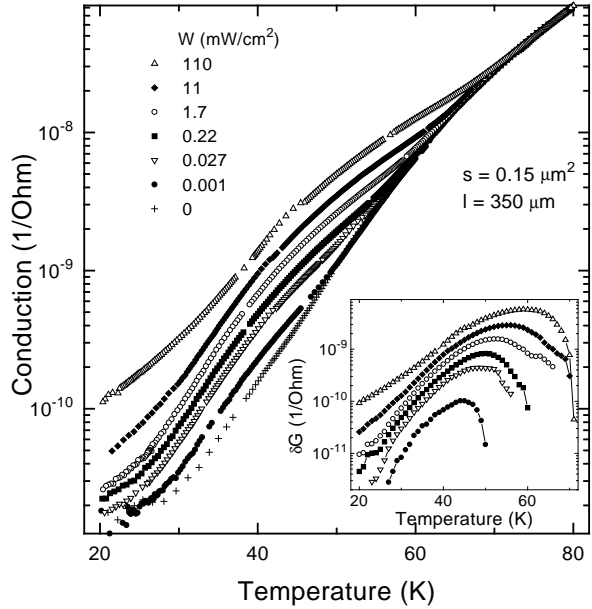}
 \vskip -2.5cm
 \caption{Temperature dependencies of conduction, $G$, at different 
steady light illumination intensities, $W$. The inset shows respective 
temperature dependencies of photoconduction, $\delta G=G(T,W)-G(T,0)$. 
All the curves were obtained upon cooling to eliminate the conduction 
hysteresis.}
\label{fig1}
 \end{figure}

Fig.~1 shows a set of temperature dependencies of 
conduction, $G(T,W)$, at various intensities $W$ of steady light
illumination for the most thick sample. 
 Noticeable deviation from the darkness curve starts at $T<70$~K. 
The inset shows the respective set the photoconduction, $\delta G=G(T,W)-G(T,0)$. 
The position of maximum of $\delta G(T)$ varies with $W$ in the range 40 - 65~K.
The deviation starts at somewhat higher temperatures for thinner samples. 
For example, for the thinnest one having  $s=0.002\;\mu$m$^2$ $\delta G/G\sim 0.01$\% was
observed at $T=100$~K. Maximum of $\delta G(T)$ at $W=110$~mW/cm$^2$ is at $T=65$~K.
The upper boundary of the light-induced heating may be estimated using 
$\delta G(T)$ at $T=100$~K to be as small as $( \delta G/G)/(dG/dT)\sim 1$~mK, 
as a consequence of exceptionally good thermal contact between the sample and substrate.
All the results presented below were obtained for this thinnest sample having the length 
$l=340\;\mu$m and the room-temperature resistance $R_{300}=430$~kOh.

\begin{figure}
 \vskip -1.8cm
 \includegraphics[scale=.48]{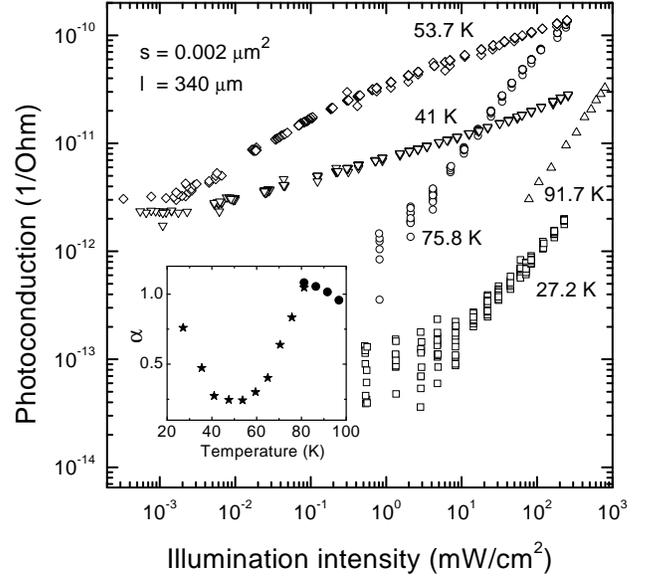}
 \vskip -4.3cm
\caption{ The dependences of photoconduction $\delta G$ on the light 
illumination intensity $W$ at different temperatures. The inset shows the 
temperature dependence of the exponent $\alpha =d\ln \delta I/d\ln W$ (different
marks correspond to the different methods of the measurements, 
see the text for detailes).}
\label{fig2}
\end{figure}

The dependence of the value of photoconduction $\delta G$ on the light
illumination intensity $W$ was studied by two methods. At the temperature range 
27 - 95~K AC conduction 
was measured at the frequency $f = 4.5$~Hz as a function of 
intensity $W$ of a steady light illumination. Such a low frequency value was 
chosen to enable measurements of very low conduction values.
At temperatures $T > 90$~K the ratio $\delta G/G$ becomes too small and 
comparable with one resulting from  temperature fluctuations. To 
improve measurements quality the  double modulation 
method was used at $T = 81$ - 100~K. Namely, the modulation of AC conduction due 
to the light illumination was detected at the frequency of light chopping 
$f_{ch} = 4.5$~Hz, AC conduction being measured at the frequency $f = 335$~Hz. 
The results are presented in Fig.~2. The dependencies can be approximated by the 
power law $\delta I = W^\alpha$, where the photocurrent $\delta I=\delta G V$. 
These dependencies are highly nonlinear in the 
middle of the temperature range ($\alpha < 1$), and approach to the linear ones 
($\alpha \approx 1$) at its ends.  The temperature dependence of the exponent 
$\alpha$ determined as $\alpha = d\ln \delta I/d\ln W$ at $W > 10$~mW/cm$^2$ is 
shown in the inset in Fig.~2. This dependence has a pronounced minimum at $T = 
50$~K where the change of the light intensity up to 4 orders of magnitude leads 
to the increase of the linear conduction only up to an order of magnitude.

\begin{figure}
  \vskip -2.2cm
 \includegraphics[scale=.45]{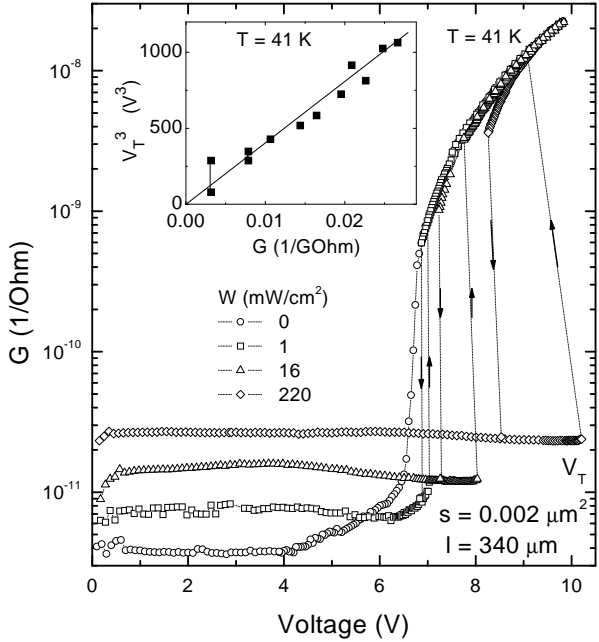}
 \vskip -2cm
\caption{The dependencies of conduction $G=I/V$ on the sample voltage $V$ at 
different steady illumination intensities $W$. The arrows show directions of the 
current sweep.The inset shows the dependence $V_T^3$ 
on linear conduction $G$ (at 100 mV) at different light intensities. 
Vertical lines correspond to uncertaincy in determination of $V_T$ 
at $W = 0$ and $W = 1$~mW/cm$^2$. }
\label{fig3}
\end{figure}

At sufficiently low temperatures the light-induced conduction variation  
becomes so large that it can be clearly seen in I-V curves. 
Fig.~3 shows the evolution of I-V curves (plotted as $G=I/V$ {\em vs.} $V$ )
caused by light illumination  at $T=41$~K. 
In the darkness  the $G(V)$ curve has the usual shape:
there is a region of a constant (linear) conductivity in a small electric field, 
then a region of creep (weak 
nonlinearity), and a region of CDW sliding (strong nonlinearity). 
The curve has a smooth character without any switching and hysteresis. 
The dramatic changes of the shape of $G(V)$ curves both in the linear 
and nonlinear conduction take place under light illumination. 
It can be seen that the growth of the light intensity causes\\
1) increase of the linear conduction up to 10 times, \\
2) decrease of the conduction with voltage growth in the creeping region, \\
3) increase of up to 60\% in the threshold field for the onset of CDW sliding  \\
(decrease of the nonlinear conduction near $V_T$), and \\
4) appearance of the switching regime (unusual for TaS$_3$ samples)
with hysteretic character \cite{Comment2}. 
The growth of the threshold field and appearance of the switching 
under light illumination was also 
observed in K$_{0.3}$MoO$_3$ \cite{Ogawa}.

Fig.~4 shows the temperature evolution of $G(V)$ dependencies 
measured under steady light illumination ($W = 117$~mW/cm$^2$) and in 
the darkness. At temperatures above 65 K (corresponding to
the maximum of the value of photoconduction for this sample) 
the changes of the shape of $G(V)$ 
curves under light illumination are visible only in the linear conduction 
region and in the region of creep. 
With temperature decrease the dependencies $G(V)$ under 
light illumination begin to deviate from the dark ones in the 
CDW sliding region as well. 
The transition to the sliding regime becomes more
sharp, the switching behavior appears, 
and the dramatic growth of the width of the hysteresis loop develops. 

\begin{figure}
\vskip -2.2cm
\includegraphics[scale=.45]{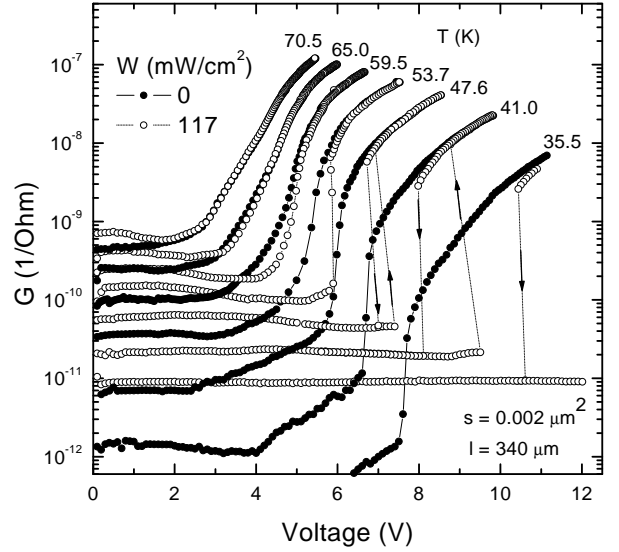}
\vskip -3cm
\caption{The dependencies of conduction $G$ on the sample voltage $V$ at
different temperatures. The dark circles correspond to the measurements in 
the darkness, the open circles correspond to the ones under steady 
light illumination.}
 \label{fig4}
\end{figure}

It is worth to note that all the $G(V)$ curves coincide in the high 
voltage region $E\gg E_T$. This indicates that 
1)  the relative variation of the nonlinear conduction due to illumination
is much less than for the linear conduction,
2) the heating of the sample due to illumination is 
small indeed (heating leads to an increase of the conduction and 
decrease of the threshold field). 

Our results allow
to estimate the recombination time $\tau$ of photoexcited carriers in TaS$_3$. 
The photoexcited current carrier concentration $\Delta n$ is 
determined by the balance between the  rate of their photogeneration, 
$kaLn_{ph}(1-e^{-\beta b})$, 
and the relaxation rate $abL\Delta n/\tau $ (we assume $\Delta n \ll n$), 
where $k$ is the quantum efficiency of photogeneration,
$n_{ph}$ is the number of incident photons per area per unit time, 
$\beta$ is the absorption coefficient, and $a$, $b$ and $L$ are the 
sample width, thickness and length, respectively. 
Thus, for $\beta b\ll 1$ one gets $\tau=\Delta n /k \beta n_{ph}$.
At $T=100$~K $\Delta n (\propto \delta G) \propto n_{ph} (\propto W)$, 
i.e. $\tau $ is independent of the light intensity $W$. As mentioned above, 
for this sample $\delta G/G \sim 0.01$\% at $T=100$~K. 
As $\Delta n/n(T)=\delta G/G(T)$, 
$n_{100}\approx 10^{-3}n_{300}$, $n_{300}\approx 2\times 10^{21}$~cm$^{-3}$, 
so $\Delta n \approx 2\times 10^{14}$~cm$^{-3}$. The value of $k$ is 
unknown. Assuming $k=1$, $1/\beta=0.3$~$\mu m$ \cite{Comment1}, for 
$W=110$~mW/cm$^2$ one gets $\tau_{100} \approx 10^{-10}$~s. Moreover, 
this value is almost two orders of magnitude greater at 65~K and 
$W=110$~mW/cm$^2$ (see Fig. 1). This time is much smaller than that in pure 
semiconductors. This explains why the photoconduction in quasi-1D  
conductors was not found during a long time. 
On the other hand, $\tau$ in thin crystals of TaS$_3$ is  
orders of magnitude greater $\tau\approx 5\times 10^{-13}$~s measured by optical 
methods in blue bronze at the same temperatures \cite{Demsar}. 
Possible physical mechanism providing enhancement of photoconduction in thin samples
is given in the following paragraph.

In the temperature range $30$~K$<T<70$~K there is a nonlinear relation between 
the light intensity and photoconduction (see inset in Fig.~2), 
and  the photoconduction reaches its maximum (inset in Fig.~1). 
These means that in this temperature range the recombination time of 
photoexcited current carriers 1) depends on their concentration, 
2) is larger than that out of this temperature range. 
These features are peculiar to 
so-called ``persistent  photoconduction''  (also ``delayed'' or ``frozen'') 
well-known for some inhomogeneous 
semiconductors \cite{persistent}. In our case the barriers for recombination result  from the 
potential relief caused by the pinned CDW. The energy range for this relief 
can be estimated from the relation
$\delta \zeta \sim E_T L_\parallel$, where  $L_\parallel$ 
is CDW phase correlation length. At higher temperatures,
$E_T$ is getting smaller (see Fig.~4), and finally one gets $\delta \zeta\ll T$. 
As a result, the relaxation barriers diminish and the relaxation is getting 
faster with a rate practically independent of the light intensity. 
As the threshold field rapidly increases with temperature decrease, 
so a situation when $\delta \zeta$ is of the order of the Peierls gap is 
achieved. In this case one may expect nucleation of dislocations and 
opening of a new relaxation channel. 
This may explain the low-temperature decrease of photoconduction 
(see inset in Fig.~1).
In addition, since $E_T$ is illumination-dependent, the barrier height depends 
on the concentration of photoexcited current carriers. Taking 
$L_\parallel=1$ - $10\;\mu$m \cite{Finite-Size} 
and $E_T \approx 30$~mV/$\mu$m for $T=41$~K at $W=220$~mW/cm$^2$ 
 (see Fig.~3) one gets 
$\delta \zeta \simeq 300$ - 3000~K$\gg T$ and 
comparable to the Peierls energy gap of TaS$_3$ (1700~K).
It is clear that such barriers have pronounced effect on the recombination rate. 
As $E_T$ in thin samples is orders of magnitude higher its value in bulk 
ones, so thin samples are preferable for observation of 
photoconduction.

Another interesting feature of I-V curves of illuminated samples 
is a decrease of conduction with growth of the voltage in the creep regime
which is clearly seen in Fig.~3 at $E\lesssim E_T$. 
From a formal point of view, this behavior corresponds to a negative 
contribution of creeping CDW to the total conduction of the illuminated sample. 
We believe, however, that this behavior results from the decrease of the 
concentration of photoexcited current carriers in the creeping regime, 
and possible contribution caused by CDW configuration variation due to  
change of current carrier concentration (configurational photoconduction). 
Such a decrease corresponds to an increase of the recombination rate 
of photoexcited current carriers due to modification and a time evolution
of the potential relief in the creeping regime. 
Note that Ogawa {\em et al.}  \cite{Ogawa} attributed
the growth of the total conduction at $E\lesssim E_T$ to the increase 
of CDW creep rate, i.e. to the opposite effect.

The illumination-induced variation of current carriers concentration must affect 
all static and dynamic properties of the CDW due to modification of the 
screening conditions. In particular, one can expect changes in CDW wave vector, 
relaxation rate of CDW metastable states, $E_T$, dielectric constant 
($\epsilon \propto 1/E_T$ \cite{Review}), CDW transport coefficients, {\em etc}.
Our results allows to verify for the first time the relations 
between the screening carrier concentration, CDW elasticity and $E_T$.
As the transverse sizes of the our samples 
are smaller than the transverse CDW phase-correlation length, 
CDW pinning is one-dimensional \cite{Finite-Size}. In this case 
$E_T\approx  (n_i w/K_\parallel)^{1/3}$, where $n_i$ is the impurity 
concentration, $w$ is the pinning potential, and $K_\parallel$ is the 
elastic modulus of the CDW \cite{Review,Finite-Size}. 
As $K_\parallel\propto 1/n$ \cite{Art}, so $E_T\propto n^{1/3}$. 
Inset in Fig.~3 shows $V_T^3$ {\em vs} $G$ dependence. 
The dependence is close to the linear one indeed.

It is well known that the energy dissipation for sliding CDW is provided 
by quasiparticles participating in screening of the time-dependent CDW 
deformations. Thus one could expect that 
growth of the current carrier concentration due to illumination would 
enhance the CDW conduction. However no noticeable modification of the CDW 
nonlinear conduction at $E\gg E_T$ is observed. Thus we conclude that 
the concentration of photoexcited current carriers diminishes with growth 
of CDW velocity. Noticeable reduction of the photoconduction 
starts already in the CDW creep region (Fig. 3).

The origin of the switching regime is a subject of wide discussions \cite{Switching}.
This regime is observed in NbSe$_3$ where the dielectrization of the electron
spectrum accompanied by CDW formation is not complete and leads to the presence 
of free carriers in the system at low temperatures. Our results prove that 
the origin of the switching effect deals with the appearance of extra current 
carriers in the system despite their nature --- natural ones as in NbSe$_3$ case 
or photoexcited ones as in TaS$_3$ case (present work) or in K$_{0.3}$MoO$_3$ 
case \cite{Ogawa}.

In conclusion, it was found that the light illumination of thin TaS$_3$ 
samples affects practically all their electrophysical properties. 
Namely, in illuminated samples we observed: \\
1) significant increase of the linear conduction; \\
2) strong nonmonotonic temperature dependence of the photoconduction;\\
3) highly nonlinear dependence of photoconduction on the light intensity;\\
4) decrease of conduction with growth of the voltage in the creep regime;\\
5) increase of the threshold field and suppression of the nonlinear conduction 
near $V_T$; \\
6) negligible effect on the nonlinear conduction at $V\gg V_T$;\\
7) appearance of the switching in I-V curves.\\
Observation of the photoresponse in the linear conduction is crucial for 
understanding the photoinduced changes in the CDW dynamics. Photoconduction in 
CDW conductors is much more complex phenomena than one in usual semiconductors 
because of very strong coupling between quasiparticles and  the CDW.
Possibility of photocontrol of quasiparticle conduction opens  wide prospects for 
investigations of various static and dynamic properties of quasi-1D conductors.

\begin{acknowledgments}
We are grateful to S.~N.~Artemenko and V.~Ya.~Pokrovskii for useful 
discussions and to R.~E.~Thorne for providing high-quality crystals.
The work was supported by the RFBR (project 04-02-16509), Presidium of RAS,
NWO, CRDF (RP2-2563-MO-03), and INTAS-01-0474.
\end{acknowledgments}



\begin{references}
\bibitem{Comment2}I-V measurements were made in 
the voltage controlled regime with a load resistance 10 MOhm. As a result, 
the current jumps were not vertical for large currents.
\end{references}
 \end{document}